%% file: egbib.tex
\documentclass[10pt,twocolumn,letterpaper]{article}

\usepackage{3dv}
\usepackage{times}
\usepackage{epsfig}
\usepackage{graphicx}
\usepackage{amsmath}
\usepackage{amssymb}

\threedvfinalcopy 
\def\threedvPaperID{223} 

\ifthreedvfinal\fi
\begin{document}

\title{gCoRF: Generative Compositional Radiance Fields}

\author{Mallikarjun BR$^{1}$~~~Ayush Tewari$^{1,2}$~~~Xingang Pan$^{1}$~~~Mohamed Elgharib$^{1}$~~~Christian Theobalt$^{1}$\\ 
		$^1$Max Planck Insitute for Informatics~~~$^2$MIT
}
\maketitle

\input{3dv/sections/abstract.tex}
\input{3dv/sections/introduction.tex}
\input{3dv/sections/related.tex} 
\input{3dv/sections/method.tex}
\input{3dv/sections/results.tex} 
\input{3dv/sections/conclusion.tex}

\let\thefootnote\relax\footnotetext{
	\textbf{Acknowledgements:}
This work was supported by the ERC Consolidator Grant 4DReply (770784).
}

{\small
\bibliographystyle{ieee_fullname}
\bibliography{egbib}
}

\end{document}

%% file: 3dv/sections/abstract.tex
\begin{abstract}

3D generative models of objects enable photorealistic image synthesis with 3D control. 
Existing methods model the scene as a global scene representation, ignoring the compositional aspect of the 
scene. 
Compositional reasoning can enable a wide variety of editing applications, in addition to enabling generalizable 3D reasoning. 
In this paper, we present a compositional generative model, where each semantic part of the object is represented as an independent 3D representation learnt from only in-the-wild 2D data. 
We start with a global generative model (GAN) and learn to decompose it into different semantic parts using supervision from 2D segmentation masks. 
We then learn to composite independently sampled parts in order to create coherent global scenes. 
Different parts can be independently sampled, while keeping rest of the object fixed.
We evaluate our method on a wide variety of objects and parts, and demonstrate editing applications. 
\end{abstract}

%% file: 3dv/sections/introduction.tex
\section{Introduction}
3D GANs are generative models capable of producing 3D representations as output that can then be rendered into RGB images. 
These models produce high-quality real images with explicit control over the camera parameters. 
While these models reason about the 3D scene as a whole, objects in the real world is made up of many different semantic components composed together. 
Compositional reasoning is thus very useful for scene understanding and synthesis. 
Compositional models allow for independent control of the different parts, while keeping rest of the sample fixed. 
For example, even though real world portrait datasets mostly contain images of male subjects with short hair, compositional reasoning of the face and hair regions allows for generalization to images of male subjects with long hair. 
While compositional models have been studied in 2D ~\cite{zhu2015learning,azadi2020compositional,lin2018st,tsai2017deep}, they have been underexplored for 3D learning.
The approach of Yang~\etal~\cite{yang2021objectnerf} learns compositional object-centric NeRF models from supervised multi-view videos of a scene.  
While this approach is scene specific, our goal instead is to learn from monocular and static image collections. 
GIRAFFE~\cite{Niemeyer2020GIRAFFE} trains a compositional 3D GAN which uses a NeRF-like volumetric representation to model the scene and separates the scene into multiple components in an unsupervised manner. 
However, GIRAFFE only models composition of multiple intact objects, but cannot model semantic parts within each object, which is our main focus in this work.

Thus, we present gCoRF, Generative Compositional Radiance Fields for synthesizing 
3D volumes of objects represented as a composition of semantic parts. 
We train our models on datasets of just monocular in-the-wild image collections.
In addition, we use an automatically computed segmentation map for each image to define the semantic parts.
Thus, unlike existing methods like GIRAFFE~\cite{Niemeyer2020GIRAFFE}, the definition of parts is well-defined in our case.
Our method learns to represent each object instance in 3D using multiple volumetric models, each representing one semantic part. %
The parts are composited using volumetric blending. 
We initialize all the part models with a global object model, and then learn to decouple it in a supervised manner. 
We also train a blending network which learns to align these sampled parts into a coherent scene, which can be rendered from any viewpoint.
At test time, each semantic part can be independently changed while keeping rest of the volume fixed.
In summary, we make the following contributions: 
\begin{enumerate}
    \item A method to learn decoupled semantic part 3D GANs of an object category using monocular image collections and corresponding semantic segmentation maps.
    \item Given decoupled part 3D GANs, we learn a blending network which can composite independently sampled parts to synthesize a coherent object volume.
    \item We show applications such as volumetric editing of various face parts (including eyes, eyebrows, nose and hair style), keeping rest of the face fixed. Here, we process both human faces and cat faces.
\end{enumerate}

%% file: 3dv/sections/related.tex
\section{Related Works}

\subsection{3D Generative Adversarial Networks}

Generative Adversarial Networks (GANs)~\cite{goodfellow2014generative} have witnessed great success in synthesizing 2D images ~\cite{Karras_2019_CVPR,karras2020analyzing,Karras2021}, but cannot fully model the 3D nature of the visual world.
Recently, to enable 3D-aware image synthesis that allows explicit viewpoint control, there is a surge of interest to develop 3D GANs.
While some works use 3D supervision~\cite{wu2016learning,chen2021decor,Gao19,yenamandra2021i3dmm}, we focus on approaches trained on monocular image collection only as images are easy to collect in large scale.
The methods~\cite{nguyen2019hologan,nguyen2020blockgan,liao2020towards} combine voxelized feature grid with 2D convolutional neural renderer for 3D-aware image synthesis.
While achieving promising results, the 3D consistency is limited.
Henzler \etal~\cite{henzler2019escaping} and Szabo \etal~\cite{szabo2019unsupervised} learn to directly generate voxels and meshes respectively, but show artifacts due to the difficulty in training.
Recently, the prosperous of NeRF~\cite{mildenhall2020nerf} has motivate researchers to use coordinate-based implicit functions as the representation for 3D GANs.
In particular, GRAF~\cite{Schwarz2020NEURIPS} and pi-GAN~\cite{piGAN2021} have shown large potential of NeRF-based GAN for 3D aware image synthesis.
They are then extended to learn more accurate 3D object shapes~\cite{pan2021shadegan,xu2021generative,or2021stylesdf,chan2021efficient} and more photorealistic image synthesis~\cite{Niemeyer2020GIRAFFE,gu2022stylenerf,or2021stylesdf,chan2021efficient}.
These advances have largely pushed forward the boundary of 3D-aware image synthesis. 

Among these works, some of them also study compositional image synthesis~\cite{Niemeyer2020GIRAFFE,nguyen2020blockgan,liao2020towards}.
All these works model composition in the object level (\eg, shifting or inserting an intact object) but cannot edit a part of an object without changing other parts.
In contrast, our method can model the compositionality within an object, allowing 3D object synthesis with more fine-grained control.
For instance, for any face volume generated by our model, we can edit one part like hair or eyes while keeping other parts fixed.

Recently proposed FENeRF~\cite{sun2021fenerf} is the closest related concurrent work to our method. While this method enables editing of parts, this requires a 2D segmentation mask for the edit. 
In contrast, our method enables one to choose any sample from the part generator. And it is also not possible to have different texture for parts using FENeRF as they have single color latent vector to define the whole face. For example, they cannot have different hair color for the edited hair.
While there are many 2D face editing methods~\cite{Tan20, saha2021LOHO, Zhu21} exist in the literature, none of the methods work without segmentation mask as input at test time with a trained model.
Also, these methods expect the segmentation mask to align well with the input image with similar pose. 
Since we model objects in 3D, we can account for these variations better.

\subsection{Compositional Scene Representation}

Many works have studied compositional model in the 2D image level~\cite{zhu2015learning,azadi2020compositional,lin2018st,tsai2017deep,johnson2018image,locatello2020object}.
Some works learn to composite images in a harmonized way~\cite{zhu2015learning,azadi2020compositional,lin2018st,tsai2017deep} while others sythesize images from given scene graphs~\cite{johnson2018image}.
Locatello \etal ~\cite{locatello2020object} proposed a slot attention architecture that can learn object composition in an unsupervised manner.
However, these 2D approaches cannot model complex occlusions in the 3D space.

Recently, several attempts have been made to model compositional scene in the 3D space using volumetric neural fields~\cite{mildenhall2020nerf}.
Specifically, Guo \etal \cite{guo2020osf} learns the NeRF for each object independently, which naturally allows composition.
%
The approaches of Ost~\etal~\cite{Ost_2021_CVPR} and Yang~\etal~\cite{yang2021objectnerf} can decomposite a scene into objects given the supervision of object masks or bounding boxes.
uORF~\cite{yu2021unsupervised} is an unsupervised approach to discover individual objects in a scene, but is only applied to synthetic data.
Wang \etal \cite{Wang21} proposed a compositional representation that models coarse structure with voxel grid and fine detials with NeRF.
Stelzner \etal \cite{stelzner2021decomposing} proposed a method which decomposes a scene into a set of NeRFs, with each NeRF corresponding to a different object. 
All these method requires either videos, RGB-D or multi-view images as supervision.
Besides, they are designed for reconstructing scenes from existing images or videos, but cannot synthesize new objects.
In this work, we propose a compositional 3D generative model that learns from only monocular images and their corresponding segmentation maps. 
Our approach not only allows editing of existing images, but also can synthesize new samples with fine-grained control over object parts.

%% file: 3dv/sections/method.tex
\section{Method}

\begin{figure*}
\centering
\includegraphics[width=\linewidth]{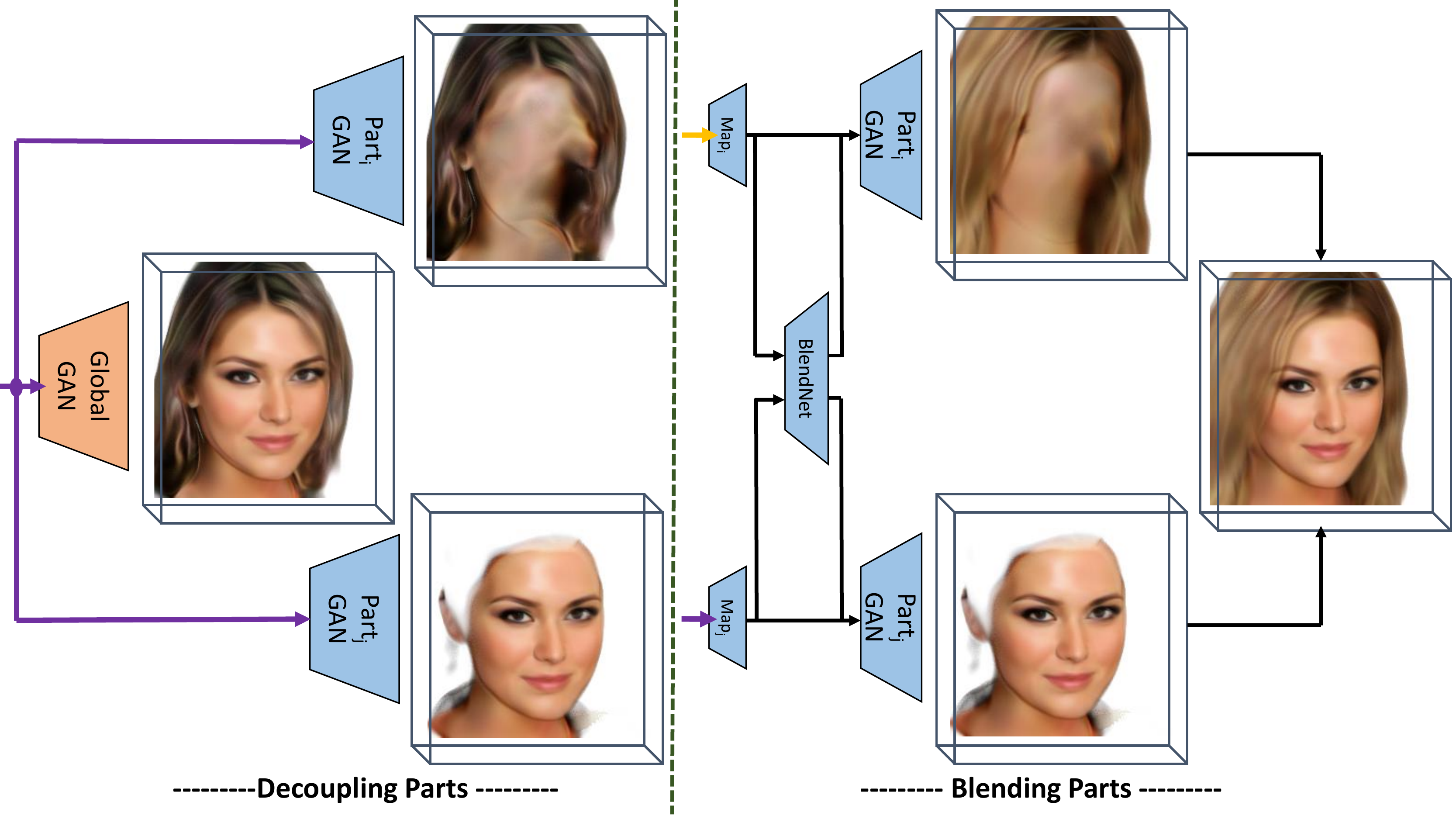}
\caption{Overview of our method. We first decompose a global GAN into several part models using supervised losses guided by segmentation masks. 
We then train a blending network which learns to composite the independently sampled parts into a coherent scene. 
This network modifies the intermediate latent vectors of the part models. 
At test time, our model enables independent sampling of the different parts in order to create photorealistic images.
}
\label{fig:pipeline}
\end{figure*}

Our method learns a compositional GAN model from just monocular in-the-wild images.
The compositional model is part-based, where each part is represented using its own MLP network as shown in Fig.~\ref{fig:pipeline}. 
This enables novel-view synthesis, as well as independent control over different parts of the object. 
Our method consists of three stages. 
First, we learn a non-compositional global-scene GAN following existing methods~\cite{piGAN2021}. 
Second, we learn to disentangle this global model into multiple models each representing a different part (Stage 1 of Fig.~\ref{fig:pipeline}). 
Finally, we learn a blending network that blends the different parts into a coherent scene for synthesis (Stage 2 of Fig.~\ref{fig:pipeline}).

\subsection{Compositional Model}
\label{sec:compositional}
Our compositional model consists of several parts, each represented using a different 3D volume. 
Thus, we have MLPs $N_i$, $i=1,\dots,P$, corresponding to $P$ parts in the object. 
The MLPs are defined as $N_i \colon (\*z_i, \*x) \to {\sigma_i, \*c_i}$. 
Here, $\*z_i$ are the latent vectors for each part, which can be independently randomly sampled from a Gaussian distribution. 
We use volumetric integration to render each part independently as $R(\*z_i, \*C)$ for a camera with parameters $\*C$. 
The output of this function is the rendered image for the part. 
For each pixel, we cast a camera ray $\mathbf{r}(t) = \mathbf{o}+ t\mathbf{d}$ with near and far bounds $t_n$ and $t_f$, origin $\mathbf{o}$ and direction $\mathbf{d}$, and compute the pixel color as:
\begin{align}
\mathbf{C}(\mathbf{r}) &= \int_{t_n}^{t_f} T(t) \sigma_i(\mathbf{r}(t)) c_i(\mathbf{r}(t)) dt \nonumber \\
\text{where} \quad & T(t) = \text{exp}(-\int_{t_n}^{t}\sigma_i(\mathbf{r}(s)) ds).
\label{eq:nerf}
\end{align}
In practice, we use a discretized version of this equation, as explained in NeRF~\cite{mildenhall2020nerf}.
A complete image of the scene can be rendered as a composition of all the individual parts. 
We first define the global volume as a composition of the individual parts, denoted with function $\mathrm{comp} \colon (\sigma_1,\dots,\sigma_P,\*c_1,\dots,\*c_P) \to (\sigma, \*c)$. 
The density $\sigma$ at each point of the global volume is a simple summation of the densities of each individual part. 
The color $\*c$ at point  $\*x$ is computed as a weighted linear interpolation of the colors $\*c_i$, where the weights are the coefficients ($T(t) \sigma_i (\mathbf{r}(t))$ for the ray $\mathbf{r}(t)$) at point $\*x$ computed in the rendering function $R(\*z_i, \*C)$, see eq.~\ref{eq:nerf}. 
These coefficients are normalized over all parts.
We can render the composite volume as $R(\mathrm{comp}(z_1,\dots,z_P, c_1,\dots,c_P), C)$.

\subsection{Decoupling Parts}
\label{sec:decoupleparts}
\paragraph{Learning a Global Model}
We use the $\pi$-GAN~\cite{piGAN2021} framework to learn the global GAN. 
Here, a single MLP network takes a randomly sampled latent code $\*z$ as input along with 3D coordinates $\*x$. 
The output of the network is a scalar density $\sigma$ and color $c \in \mathbb{R}^3$ at point $\mathbf{x}$ of object instance $\mathbf{z}$.
An image can be rendered from this 3D representation using volumetric rendering.
During training, images from the model are synthesized from a random distribution of cameras around the object. 
The network is trained in a generative adversarial framework using a discriminator, and a non-saturating adversarial loss~\cite{piGAN2021}.
\begin{align}
\mathcal{L_\text{adv}} &= f \Big( D(R (\*z, C) \Big) 
&+ f(-D(\mathbf{I})) + \lambda \|\nabla D(\mathbf{I})\|^2 .
\label{eq:gan}
\end{align}
Here, $R(\cdot)$ 
is the rendering function as explained in eq.~\ref{eq:nerf}, D is the discriminator, 
$\mathbf{I}$ are real images sampled from the training dataset, $f(u) = -\log(1 + \exp(-u))$, and $\lambda$ is the coefficient for $R_1$ regularization.
\paragraph{Generator Network}
The generator networks in our case are parameterized as MLPs. 
The randomly sampled latent vector $\*z$ is first mapped to an intermediate space $\*w$ using a ReLU MLP, following~\cite{Karras_2019_CVPR,piGAN2021}.
We use a SIREN MLP with FiLM conditioning on $\*w$ and linear conditioning on the input points to compute the density and color values at the point.
The same design is used for all part-based networks, where $\*z_i$ is first mapped into $\*w_i$.

\paragraph{Training} The second stage of our training is supervised on segmentation maps for the different parts, computed using off-the-shelf segmentation models~\cite{Yu-ECCV-BiSeNet-2018,zhang21}. 
The weights of each part network $N_i$ is initialized with the pretrained global network $N$.  
The mapping network that maps $\*z_i$ to $\*w_i$ for each part $i$ is fixed during training, and only the SIREN MLP is finetuned. 
We sample a dataset of latents $\*z$, map them into $\*w$, and compute images $\*I$ from the global model using randomly distributed cameras. 
Additionally, we compute segmentation maps $\*S_i$, $i=1,\dots,P$, where $\*S_i$ is one for pixels where part $i$ is present in the image $\*I$, and zero elsewhere.
Then, we fix $\*z_i = \*z$, $\forall i$.
The goal of this stage is to learn to separate the global model into the different parts.
We use the following loss function for training:
\begin{align*}
    \mathcal{L}_\text{part} &= \sum_{i=0}^P || \*S_i \odot (R_i(\*z, \*C) - \*I) ||^2  \\
    &+ \sum_{i=0}^P || \*S_i \odot (R(\mathrm{comp}(\*z,\dots,\*z, \*c'_1,\dots,\*c'_P), \*C) - \*I) ||^2 \\ &+  || R(\mathrm{comp}(\*z,\dots,\*z, \*c_1,\dots,\*c_P), \*C) - \*I) ||^2\,.
\end{align*}
Here, the first term compares the image difference of the rendered parts to the ground truth image in the segmented region for the part. %
The second term also compares each part to the image, while each part is rendered to also take the occlusions from other parts into account. 
This is done by setting the color of all other parts to white, i.e., for part $i$, $\*c_j' = \*1$, $\forall j \not= i$, and $\*c_i' = \*c_i$.
This encourages the density predictions for the other parts to be zero.
The third term evaluates the rendering of the composited volume with the complete input image. 
This training strategy allows us to decouple the different parts of the object in a supervised manner. 
We use hyperparameters to balance the different terms. %
Please refer to supplementary document for ablative analysis of each loss term.

\subsection{Blending Network}
\label{sec:blendnetwork}
The decoupling stage allows us to train the individual part generative models. 
However, so far, the networks were trained with coupled latent vectors, i.e., identical latent vectors for each part. 
Independently sampling the latent vectors at test time can lead to artifacts at this point. 
For example, for a model with independent face region and hair parts, the face shape learned in the second stage could be dependent on the hairstyle. %
To enable independent sampling with photorealistic output, we introduce a blending network as $B \colon  (\*w_1,\dots,\*w_P) \to (\*d_1,\dots,\*d_P)$.
The output vectors are computed as $\*w'_i = \*w_i + \*d_i, \forall i$.
This network is parameterized as a ReLU MLP. 
The goal is to update the latent vectors of the individual parts in order to compose a realistic and coherent global volume. 
We use the following loss function to train the blending network:
\begin{align*}
    \mathcal{L}_\text{blend} &= \mathcal{L}_\text{adv} + \sum_{i=0}^P || \*d_i ||^2  \\
     &+  || R_w(\mathrm{comp}_w(\*w_1',\dots,\*w_P', \*c_1,\dots,\*c_P), \*C) - \\   & R_w(\mathrm{comp}_w(\*w_1,\dots,\*w_P, \*c_1,\dots,\*c_P), \*C)) ||^2\,.
\end{align*}
Here, $\mathcal{L}_\text{adv}$ is the adversarial loss introduced in eq.~\ref{eq:gan}, and the second term encourages the latent differences to be small.
The third term is similar to the second term, where $R_w (\cdot)$ and $\textrm{comp}_w (\cdot)$ are the equivalent functions of $R(\cdot)$ and $\textrm{comp}(\cdot)$ where the input latents are the output of the mapping network. 
This term encourages the blending network to not change the latents too drastically. 
We use hyperparameters to balance the different terms. %
With this, our method enables independent sampling of the different parts of an object while being able to generate coherent volumes. 

%% file: 3dv/sections/results.tex
\section{Results}
We evaluate our method on human portraits~\cite{karras2017progressive} and cats~\cite{zhang2008cat} datasets.  
We model different semantic parts of the portraits like hair, eyes, eyebrows and nose using our method.
We used BiSeNet~\cite{Yu-ECCV-BiSeNet-2018} trained on CelebAHQMask~\cite{CelebAMask-HQ} dataset to obtain semantic map for portrait images, and  DatasetGAN~\cite{zhang21} to obtain semantic map for Cats category.
We compare to several baselines and related approaches, both qualitatively and quantitatively, and demonstrate the advantages of our method.
We provide detailed explanation of network architecture, training curriculum, hyperparamters in the supplementary document.
As our method requires separate Generator backbone for each part, because of memory and computational expense, we train our networks at $64 \times 64$ resolution, following the settings of $\pi$-GAN~\cite{piGAN2021}.
All results are rendered at $128 \times 128$ resolution. 
Please refer to supplementary document for more results and ablative analysis.

\subsection{Datasets}
We use the CelebAHQ~\cite{karras2017progressive} dataset for portrait images and Cats dataset~\cite{zhang2008cat} for cat faces.
Human portrait images are very suitable for the proposed task of 
compositional learning with semantically well-defined parts such as hair, eyes, eyebrows and nose. 
For cat faces, we show how we can learn independent models of cat eyes and the rest of the face.
As our main objective is to learn independent models for each of the face parts, we do not model the background and train on foreground segmented images.

\subsection{Qualitative Results}
\begin{figure}
\centering
\begin{minipage}[t]{.48\textwidth}
  \centering
  \includegraphics[width=\linewidth]{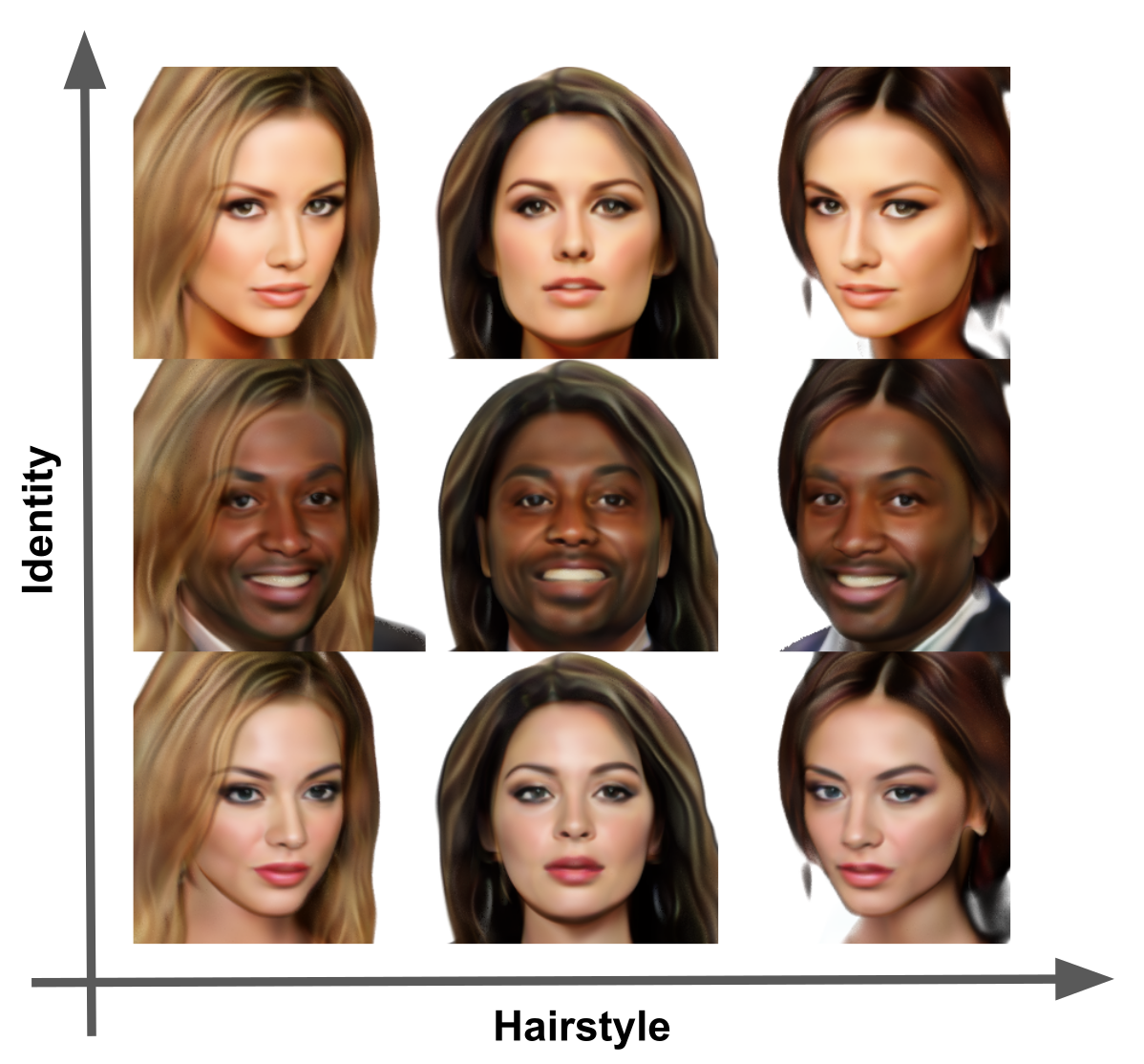}
  \caption{Independent sampling of hair and face region. Each row shows a fixed face region sample, while each column shows a fixed hair sample. We can independently sample from these models to create photorealistic images.  
  We also have explicit control over the head pose because of the underlying 3D representation. 
  }
  \label{fig:result_hair}
\end{minipage}%
\hspace{0.1cm}
\begin{minipage}[t]{.48\textwidth}
  \centering
  \includegraphics[width=\linewidth]{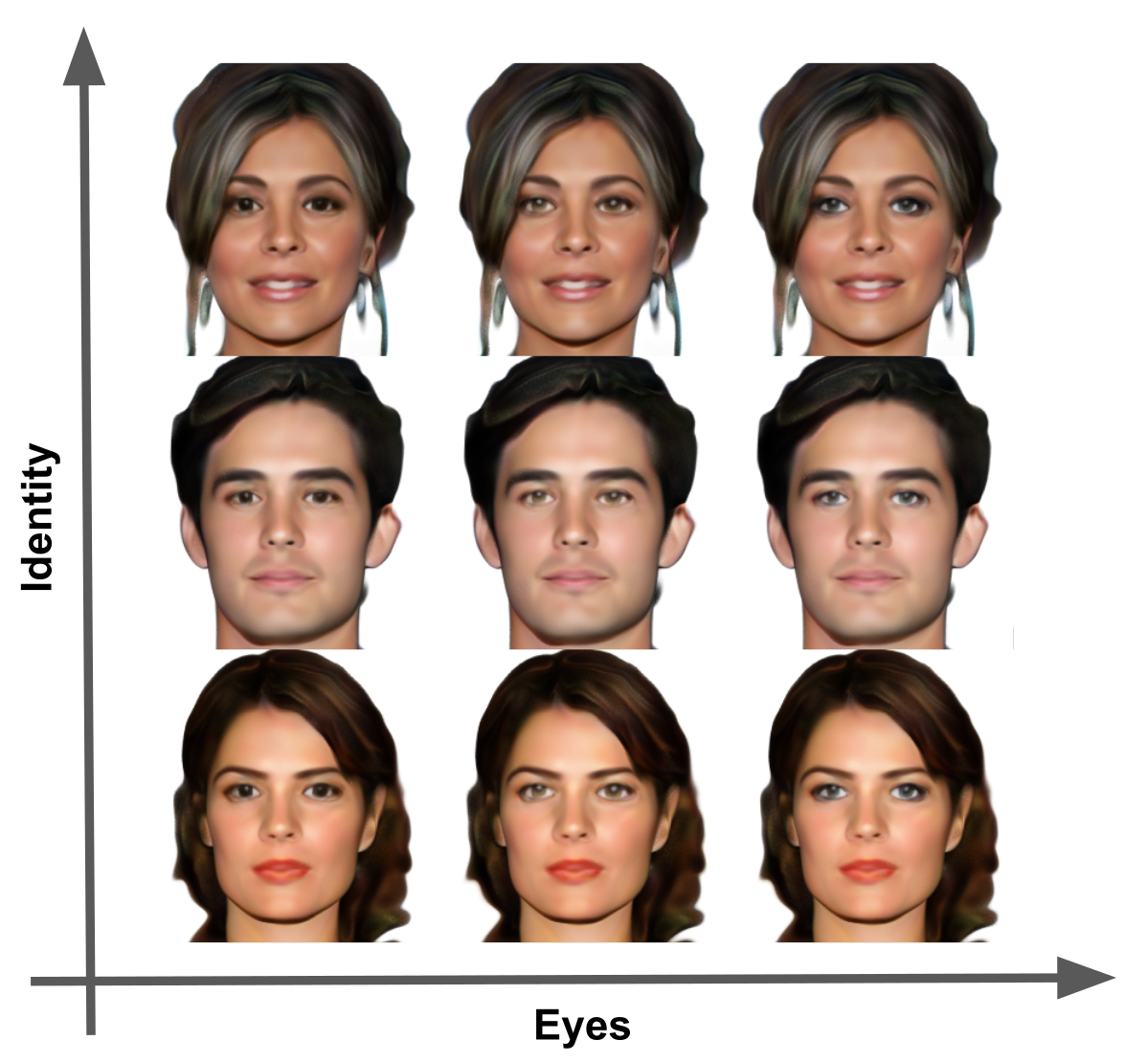}
  \caption{Independent sampling of eyes and the rest of the person. Each row shows a fixed non-eyes sample, while each column shows a fixed eyes sample. We can independently sample from these models to create photorealistic images.  
  }
  \label{fig:result_eyes}
\end{minipage}%
\end{figure}

\begin{figure}
\centering
\begin{minipage}{.48\textwidth}
  \centering
  \includegraphics[width=\linewidth]{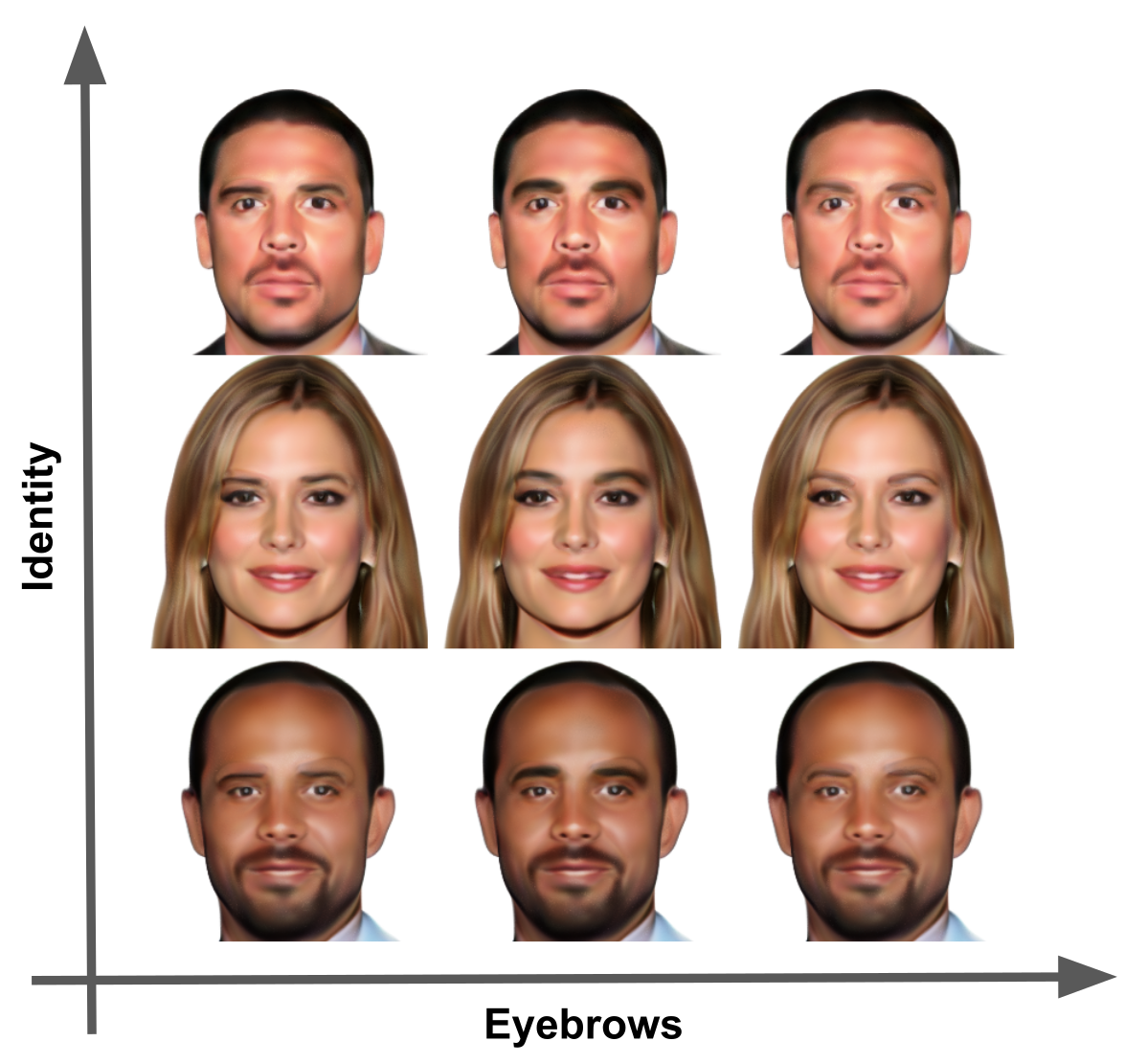}
  \caption{Independent sampling of eyebrows and the non-eyebrow region. Each row shows a fixed non-eyebrow sample, while each column shows a fixed eyebrows sample. We can independently sample from these models to create photorealistic images.  }
  \label{fig:result_eyebrows}
\end{minipage}%
\hspace{0.1cm}
\begin{minipage}{.48\textwidth}
  \centering
  \includegraphics[width=\linewidth]{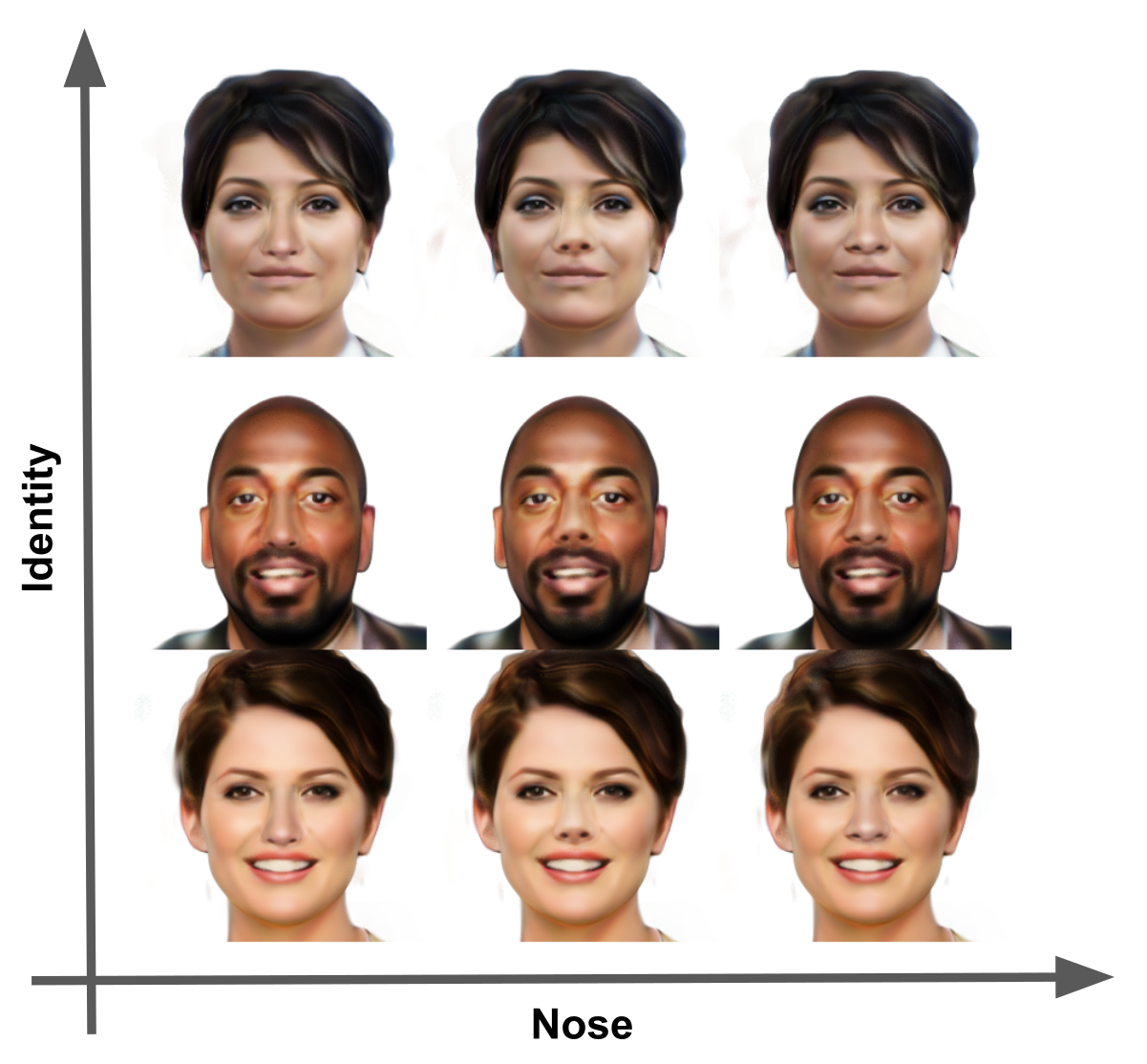}
  \caption{Independent sampling of nose and the non-nose region. Each row shows a fixed non-nose sample, while each column shows a fixed nose sample. We can independently sample from these models to create photorealistic images.  }
  \label{fig:result_nose}
\end{minipage}%
\end{figure}

\begin{figure}
\centering
\centering
\includegraphics[width=\linewidth]{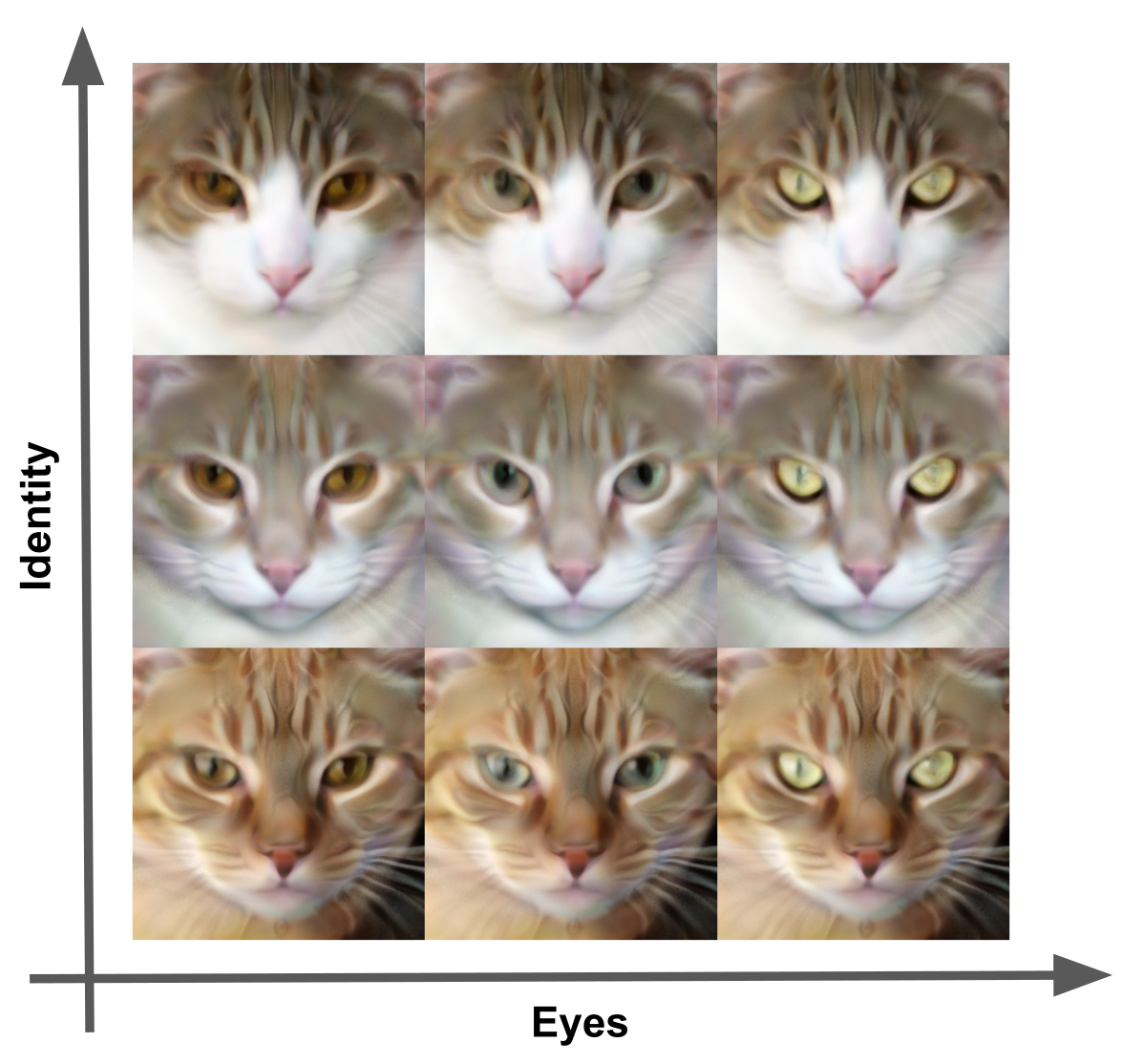}
  \caption{Independent sampling of eyes and the non-eyes region. Each row shows a fixed non-eyes sample, while each column shows a fixed eyes sample. We can independently sample from these models to create photorealistic images.  }
  \label{fig:results_cats}
\end{figure}

We show results of independent hair and face region models for portrait images in Fig.~\ref{fig:result_hair}. 
Each row shows samples of the model with the same face but different hair in different poses. 
Each column shows samples with the same hair but different faces in a fixed pose. 
Please note the diversity of hairstyles we can synthesize for any face.
We show disentangled eye and the rest of the person in Fig.~\ref{fig:result_eyes}, eyebrows and the rest of the person in Fig.~\ref{fig:result_eyebrows}, and nose and the rest of the person in Fig.~\ref{fig:result_nose}. 
Further, we show results of eyes vs rest of the scene disentanglement for cats in Fig.~\ref{fig:results_cats}.
These results demonstrate that our method works across different parts with different levels of occlusions and sizes. 
Note that we have explicit control over the head pose as our method produces a 3D representation as output. %
We request the reader to check our supplemental video, where we show renderings of smooth interpolation of view points, part samples for many cases.

Control over the different parts of the object can be very useful for various applications ranging from image editing to image analysis. 
For example, a facial recognition system trained on images could learn to distinguish between male and female subjects by looking at the hairstyle. 
Our method could potentially enable data augmentations which, while being photorealistic, ensures a balanced distribution of hairstyles for training regardless of gender. 
Please note that the contributions of this paper are to enable a rich set of control in generative models. We leave explorations in downstream tasks for future work. 

\paragraph{Comparisons to Baselines}
\begin{figure}
\centering
\includegraphics[width=0.95\linewidth]{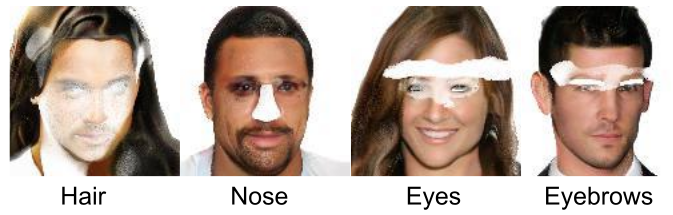}
\caption{Results of baseline with the models for each part trained independently from each other. Without the constraints between parts, such a baseline leads to artifacts due to incorrect separation.}
\label{fig:baseline}
\end{figure}

\begin{figure}
\centering
\includegraphics[width=\linewidth]{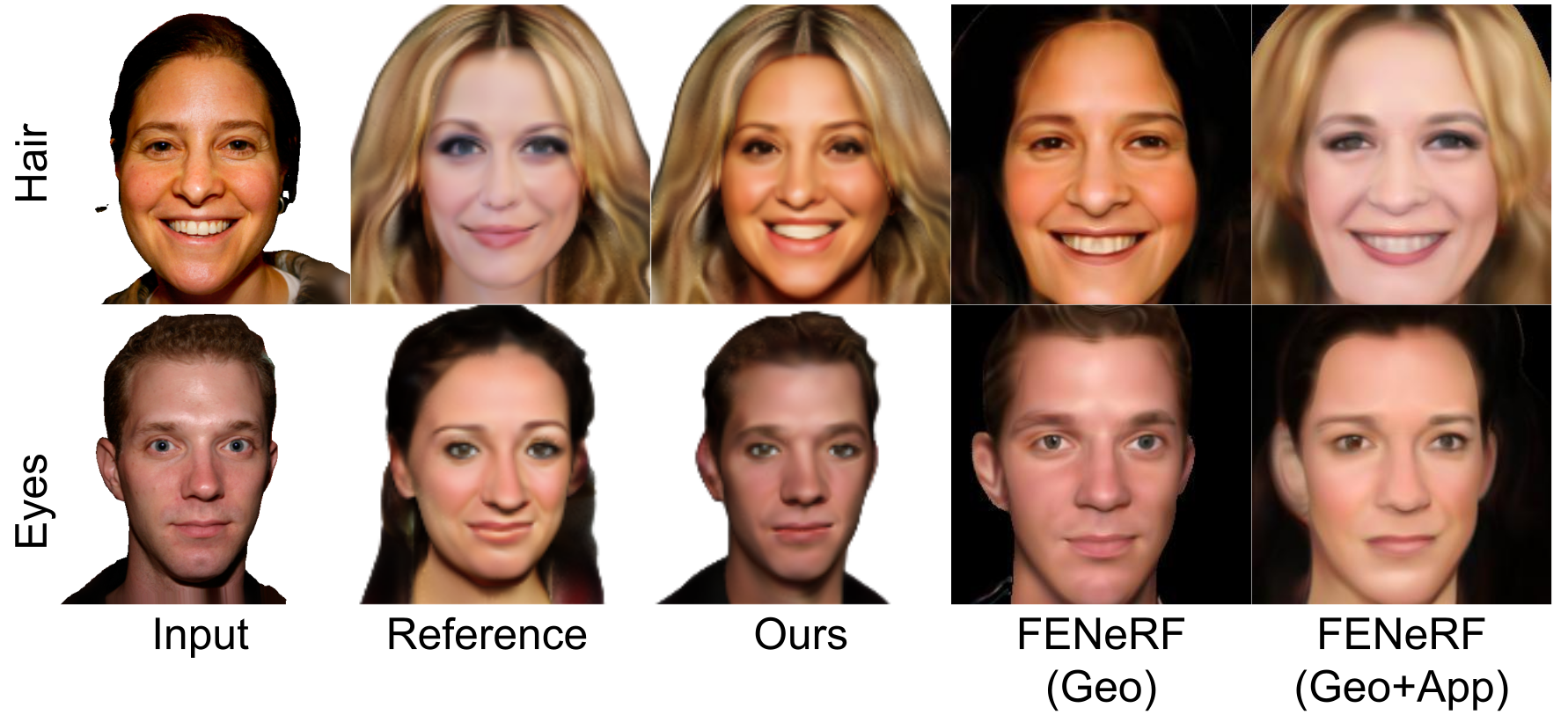}
\caption{Here we show editing results in comparison with FENeRF~\cite{sun2021fenerf}. Our method can edit not just the shape of the part, but also color independently of other parts. In contrast, FENeRF can edit the shape of the part, while keeping the part's color fixed (Geo). Editing the part color also changes the color of the whole face, as it relies on single appearance latent vector for the whole face (Geo+App). Here, "Geo" refers to "Geometry" and "App" refers to "Appearance".}
\label{fig:fenerf_comp}
\end{figure}

\begin{figure}
\centering
\includegraphics[width=\linewidth]{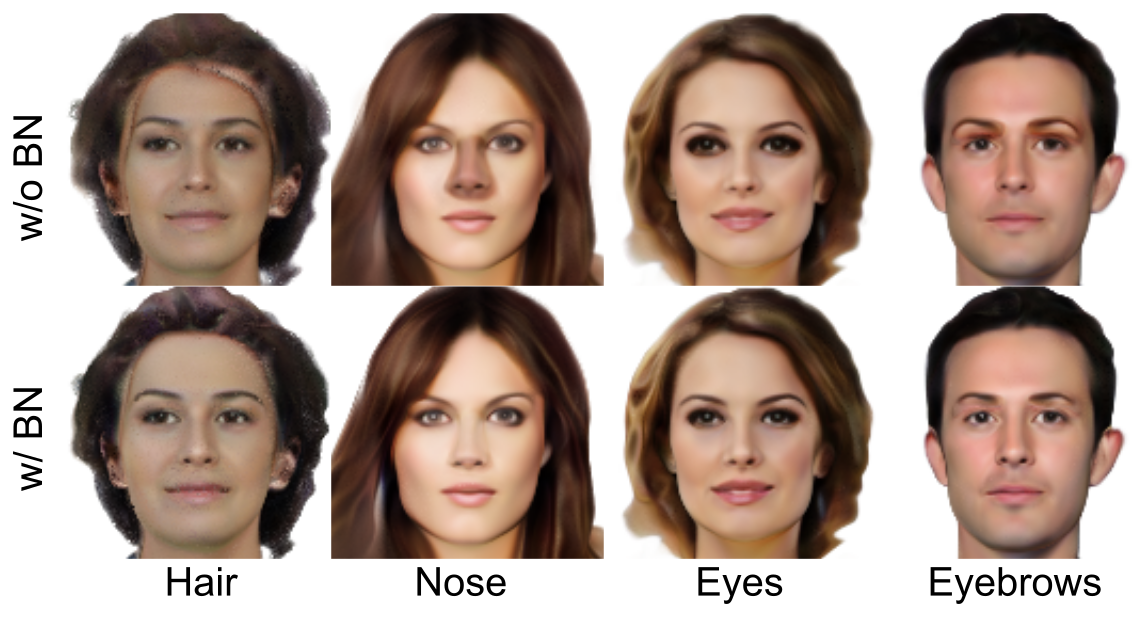}
\caption{Top row shows results without using the blending network. Without the blending network, the rendered results can have artifacts around the boundary between the parts.}
\label{fig:ablationblend}
\end{figure}

\begin{table*}[]
\centering
\begin{tabular}{llllll}
\hline
Metric    & $\pi$-GAN      &    Hair         & Eyes        & Eyebrows    & Nose        \\ \hline
FID $\downarrow$       & 16.246         &    14.8738      & 14.0873     & 13.9845     & 14.2577     \\ 
KID $\downarrow$       & 0.012          &    0.0095       & 0.0091      &  0.0093     & 0.0094      \\ 
IS  $\uparrow$      & 2.479          &    2.5371       & 2.5068      & 2.4455      & 2.4778      \\ 
\end{tabular}
\caption{
Here we show quantitative comparison with $\pi$-GAN for portrait datset~\cite{karras2017progressive}. To have fair comparison, we tie the sampling vectors for all the parts to synthesize images for evaluation, as $\pi$-GAN is not capable of synthesizing different semantic parts.}
\label{tab:coupled_sampling}
\end{table*}

\begin{table*}[]
\centering
\begin{tabular}{lllllllllll}
\hline
                       & $\pi$-GAN &   Hair     &  Eyes      &  Eyebrows          & Nose        \\  \hline

                       &           &            &  $\pi$-GAN-ind   &                   &           \\  \hline
FID $\downarrow$       & x         &   143.184  &  87.6181    &  52.924           & 86.131    \\ 
KID $\downarrow$       & x         &   0.14530  &  0.10107    &  0.0580           & 0.0981    \\
IS  $\uparrow$         & x         &   3.689    &  2.19169    &  2.4125           & 2.334     \\  \hline

          &           &            & Ours(w/o BN)   &           &            \\  \hline
FID $\downarrow$      & x         & 20.8340    & 15.2308         & 15.1184   & 18.2041    \\ 
KID $\downarrow$      & x         & 0.0145     & 0.0100          & 0.0099    & 0.01284    \\
IS  $\uparrow$      & x         & 2.1991     & 2.3435          & 2.3080    & 2.3069     \\  \hline

          & x           &           &  Ours     &             &             \\  \hline
FID $\downarrow$       & x           & 18.3422   & 13.9916   & 13.9293     & 15.1478     \\ 
KID $\downarrow$      & x           & 0.01280   & 0.0092    & 0.0093      & 0.0102      \\
IS  $\uparrow$      & x           & 2.3293    & 2.4680    & 2.4069      & 2.4436      \\  \hline

\\ 
\end{tabular}
\caption{Here we show quantitative comparison with baseline ($\pi$-GAN-ind) and ablative study of blending network for portrait datset~\cite{karras2017progressive}.  As the baseline models do not explicitly account for occlusion by other parts in the object, they end up learning some parts of the rest of the object with segmented color. As a result, the composited volume of different parts result in artifacts, which are reflected in bad numbers. }
\label{tab:random_sampling}
\end{table*}

\begin{table*}[]
\centering
\begin{tabular}{lllllll}
\hline
Metric    & $\pi$-GAN   & Ours (tied)   & \vline    $\pi$-GAN-ind    &   Ours(w/o BN)     & Ours       \\ \hline
FID $\downarrow$       & 13.3387     & 13.6265       & \vline    47.384      &   16.0410          & 14.6521               \\ 
KID $\downarrow$       & 0.0079      & 0.0066        & \vline    0.0420      &   0.0084           & 0.0075              \\ 
IS $\uparrow$       & 2.0481      & 2.0354        & \vline    2.0534      &   2.0537           & 2.0457        \\    
\end{tabular}
\caption{Here we show results for Cats dataset. The first two quantitative columns (before vertical line) show results for samples with tiedt latent vectors between parts. The last 3 columns show numbers for samples with independent part sampling.} 
\label{tab:cats_quant}
\end{table*}

\begin{figure*}
\centering
\includegraphics[width=0.8\linewidth]{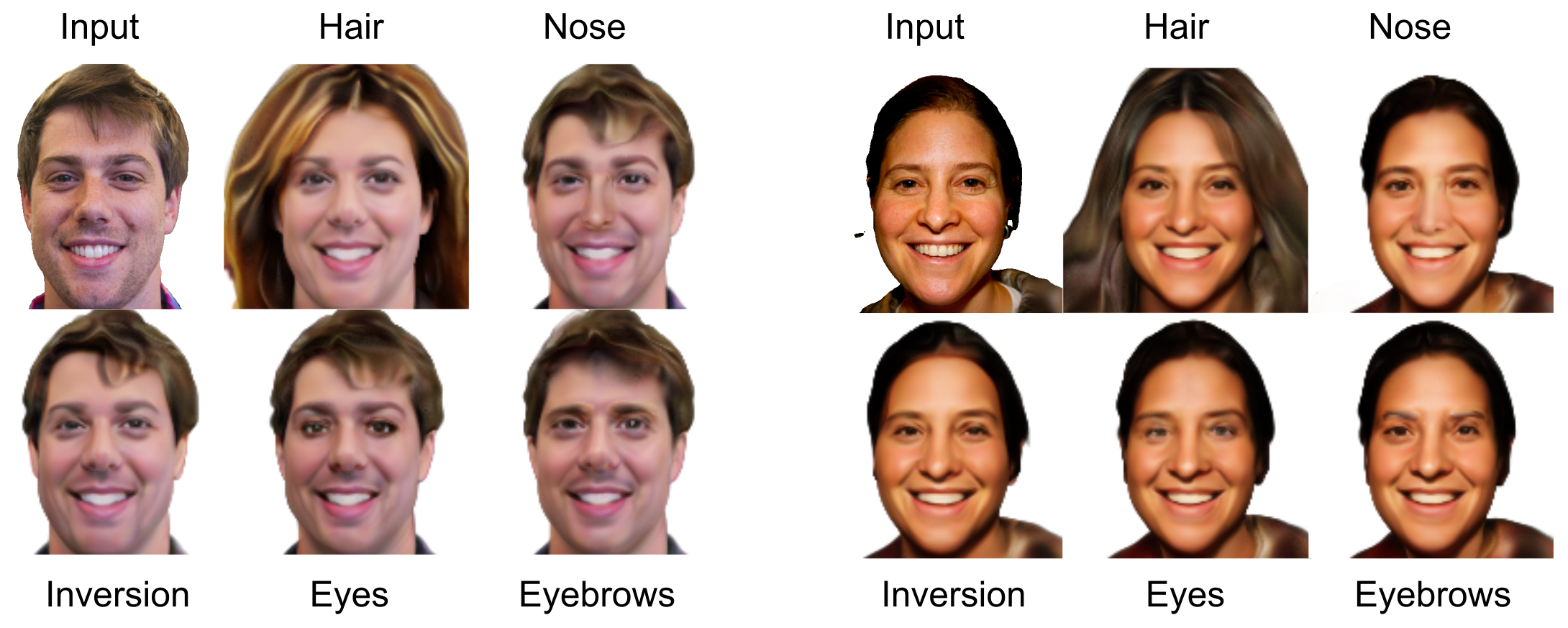}
\caption{Results on real images. We embed the input images into the latent space of our model. This enables editing of semantic parts by sampling different components independently. }
\label{fig:result_real_edit}
\end{figure*}

\begin{figure}
\centering
\includegraphics[width=0.55\linewidth]{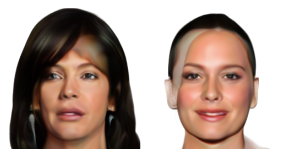}
\caption{Limitations of our method. In some cases, the independently sampled parts can be very different, which introduces artifacts in the global rendering. }
\label{fig:limitations}
\end{figure}

We design a baseline where each part model is independently trained from scratch, i.e., we train a $\pi$-GAN model for each part on the corresponding segmented part images.
Once trained, we define a model which simply composite the independently sampled volumes using the compositional operation defined in Section~\ref{sec:compositional}, denoted as $\pi$-GAN-ind.
Fig.~\ref{fig:baseline} shows the results of the composited volumes. 
This baseline does not learn a reasonable volume, as the constraints between the different parts are not modeled.
For example, the hair model can learn arbitrary depth for the face region with the color of the background. 
This geometry can be in front of the sampled face region, and can thus occlude the face. 
Our design models occlusions between the different parts and thus does not face this limitation. 

We also compare to a baseline without using a blending network in Fig.~\ref{fig:ablationblend}.
We note that, blending network helps in synthesizing texture of the part similar to that of rest of the face, while keeping the shape of the sampled part (Nose column in Fig.~\ref{fig:ablationblend}).
Without the blending network (top row), there can often be artifacts in the boundary between the different parts (check the artifact between hair and face region in first row, first column in Fig.~\ref{fig:ablationblend} ). 

\paragraph{Comparison to FENeRF}
We show qualitative comparison with FENeRF~\cite{sun2021fenerf} in Fig~\ref{fig:fenerf_comp}.
For a given input image, we can edit both the shape and color of the part independent of rest of the face. In the Fig~\ref{fig:fenerf_comp}, we take the fit our model to input image and take the corresponding part latent code from reference image to edit individual parts. FENeRF can only edit the shape while keeping the part color fixed. Editing the color of the part also changes the color of the rest of the face as they have only a global appearance vector.

\subsection{Quantitative Results}
We evaluate our model quantitatively using FID, KID, and IS scores, commonly used to evaluate generative models. 
In Table ~\ref{tab:coupled_sampling}, we compare against  $\pi$-GAN on both face and cats datasets. 
Since $\pi$-GAN does not model independent sampling of parts, to have a fair comparison to our method, we tie the latent vectors (i.e., latents for all parts are identical) when we randomly sample images. 
This evaluation also show that our training (Sec.~\ref{sec:decoupleparts}, Sec.~\ref{sec:blendnetwork}) does not negatively impact the quality of the model. 
Our decoupled model performs almost similar or slightly better to $\pi$-GAN for all categories. 
As the baseline models ($\pi$-GAN-ind) are trained independently, it can not be used for this evaluation.
In Table ~\ref{tab:random_sampling},~\ref{tab:cats_quant}, we compare our method against the $\pi$-GAN-ind baseline. In this table, we do not tie the latent vectors of the parts, but rather independently sample the latents. 
As can be seen qualitatively as well in Fig.~\ref{fig:baseline}, naively training independent part models can not be used to composite multiple parts together. 
Even though the missing parts in the image set is segmented with white color (same as back-ground), the network accounts it as part of the model and paints that region the segmented color.
Therefore, when composited with other parts, using our composite operator, it results in such white artifacts.
This can be observed quantitatively for all the categories.
Note that, the quantitative results for our method with random sampling and coupled sampling are in the similar range.
In Table ~\ref{tab:random_sampling},~\ref{tab:cats_quant}, we provide ablative analysis of our blending network.
In our experiments, we used blending network for all part categories, including hair, nose, eyes and eyebrows.
The blending network helps in overcoming certain artifacts pertaining to texture harmonization and artifacts near the boundary of composited parts in some cases, leading to better quantitative scores as verified in Table~\ref{tab:random_sampling},~\ref{tab:cats_quant}.

\section{Application - Editing Real Images}
%
Our model can enable many exciting applications. 
Here, we demonstrate the application of editing real images. 
This requires us to embed the real images into the latent space of the learned GAN. 
Projecting real images onto the latent space of GANs is a well-studied problem~\cite{xia2021gan}.
Methods either use optimization-based techniques~\cite{tewari2020pie,abdal2019image2stylegan}, or  learn an encoder to regress the latent vectors~\cite{alaluf2021restyle,richardson2021encoding}.
We design an optimization-based technique, since it tends to result in higher-quality images, however, at the cost of the speed of projection. 
We use several energy terms to optimize for the latent vector. 
We use a combination of L1 loss, perceptual loss, and facial recognition loss on the complete image. 
In addition, we use an L1 loss between the renderings of each part with the corresponding segmented out part of the input image.
Finally, we use regularizers encouraging the optimized latent vectors to be close to their average values via L2 losses. 
Please refer to the supplemental for more details on the optimization and for all hyperparameters used.
The real image editing results are shown in Fig.~\ref{fig:result_real_edit}, where several semantic parts of the portrait images are edited independently.

\section{Limitations}
While we demonstrated high-quality results for our compositional scene representation, several limitations remain. 
Fig.~\ref{fig:limitations} show that when composing independently sampled hair and face regions which are quite apart, our methods sometimes struggles to synthesize coherent volume at the boundaries.
Similar to existing 3D GANs, we can only learn our models at relatively low resolutions. %
Concurrent work has shown ways to improve the resolution and quality of 3D GANs~\cite{gu2022stylenerf,chan2021efficient}, and our framework could potentially be used with newer 3D GANs. This is left this exploration for future work.
Although our method works well for the object categories as shown in this paper, we believe it might not hold true for other categories like articulated objects (For example, human bodies, hands etc.). But our method could be extended by explicitly accounting for geometric articulation of the parts, as we can already decouple part in 3D explicitly. We leave this for future work.

%% file: 3dv/sections/conclusion.tex
\section{Conclusion}
We present a compositional 3D generative model of objects, where different semantically meaningful parts of objects are represented as independent 3D models. 
Different parts can be independently sampled to create photorealistic images. 
The disentanglement of a global model into parts is achieved using losses in image space, defined with the help of segmentation masks. 
Unlike many existing approaches, our parts are semantically well-defined. 
We validate our method using several experiments on portrait images and cat face images, and compare to several baseline and competing methods. 
Compositional models have a lot of applications in Computer Vision and Computer Graphics, and we hope that our method inspires follow-up work in this area.

%% file: egbib.bbl
\begin{thebibliography}{10}\itemsep=-1pt

\bibitem{abdal2019image2stylegan}
Rameen Abdal, Yipeng Qin, and Peter Wonka.
\newblock Image2stylegan: How to embed images into the stylegan latent space?
\newblock In {\em Proceedings of the IEEE/CVF International Conference on
  Computer Vision}, pages 4432--4441, 2019.

\bibitem{alaluf2021restyle}
Yuval Alaluf, Or Patashnik, and Daniel Cohen-Or.
\newblock Restyle: A residual-based stylegan encoder via iterative refinement.
\newblock In {\em Proceedings of the IEEE/CVF International Conference on
  Computer Vision (ICCV)}, October 2021.

\bibitem{azadi2020compositional}
Samaneh Azadi, Deepak Pathak, Sayna Ebrahimi, and Trevor Darrell.
\newblock Compositional gan: Learning image-conditional binary composition.
\newblock {\em International Journal of Computer Vision}, 128(10):2570--2585,
  2020.

\bibitem{piGAN2021}
Eric Chan, Marco Monteiro, Petr Kellnhofer, Jiajun Wu, and Gordon Wetzstein.
\newblock pi-gan: Periodic implicit generative adversarial networks for
  3d-aware image synthesis.
\newblock In {\em Proc. CVPR}, 2021.

\bibitem{chan2021efficient}
Eric~R Chan, Connor~Z Lin, Matthew~A Chan, Koki Nagano, Boxiao Pan, Shalini
  De~Mello, Orazio Gallo, Leonidas Guibas, Jonathan Tremblay, Sameh Khamis,
  et~al.
\newblock Efficient geometry-aware 3d generative adversarial networks.
\newblock {\em arXiv preprint arXiv:2112.07945}, 2021.

\bibitem{chen2021decor}
Zhiqin Chen, Vladimir~G Kim, Matthew Fisher, Noam Aigerman, Hao Zhang, and
  Siddhartha Chaudhuri.
\newblock Decor-gan: 3d shape detailization by conditional refinement.
\newblock In {\em Proceedings of the IEEE/CVF Conference on Computer Vision and
  Pattern Recognition}, pages 15740--15749, 2021.

\bibitem{Gao19}
Lin Gao, Jie Yang, Tong Wu, Yu-Jie Yuan, Hongbo Fu, Yu-Kun Lai, and Hao Zhang.
\newblock Sdm-net: Deep generative network for structured deformable mesh.
\newblock {\em ACM Trans. Graph.}, 38(6), nov 2019.

\bibitem{goodfellow2014generative}
Ian Goodfellow, Jean Pouget-Abadie, Mehdi Mirza, Bing Xu, David Warde-Farley,
  Sherjil Ozair, Aaron Courville, and Yoshua Bengio.
\newblock Generative adversarial nets.
\newblock In {\em NIPS}, 2014.

\bibitem{gu2022stylenerf}
Jiatao Gu, Lingjie Liu, Peng Wang, and Christian Theobalt.
\newblock Stylenerf: A style-based 3d-aware generator for high-resolution image
  synthesis.
\newblock 2022.

\bibitem{guo2020osf}
Michelle Guo, Alireza Fathi, Jiajun Wu, and Thomas Funkhouser.
\newblock Object-centric neural scene rendering.
\newblock {\em arXiv preprint arXiv:2012.08503}, 2020.

\bibitem{henzler2019escaping}
Philipp Henzler, Niloy~J Mitra, and Tobias Ritschel.
\newblock Escaping plato's cave: 3d shape from adversarial rendering.
\newblock In {\em ICCV}, 2019.

\bibitem{johnson2018image}
Justin Johnson, Agrim Gupta, and Li Fei-Fei.
\newblock Image generation from scene graphs.
\newblock In {\em Proceedings of the IEEE conference on computer vision and
  pattern recognition}, pages 1219--1228, 2018.

\bibitem{karras2017progressive}
Tero Karras, Timo Aila, Samuli Laine, and Jaakko Lehtinen.
\newblock Progressive growing of gans for improved quality, stability, and
  variation.
\newblock {\em arXiv preprint arXiv:1710.10196}, 2017.

\bibitem{Karras2021}
Tero Karras, Miika Aittala, Samuli Laine, Erik H\"ark\"onen, Janne Hellsten,
  Jaakko Lehtinen, and Timo Aila.
\newblock Alias-free generative adversarial networks.
\newblock In {\em Proc. NeurIPS}, 2021.

\bibitem{Karras_2019_CVPR}
Tero Karras, Samuli Laine, and Timo Aila.
\newblock A style-based generator architecture for generative adversarial
  networks.
\newblock In {\em Proceedings of the IEEE/CVF Conference on Computer Vision and
  Pattern Recognition (CVPR)}, June 2019.

\bibitem{karras2020analyzing}
Tero Karras, Samuli Laine, Miika Aittala, Janne Hellsten, Jaakko Lehtinen, and
  Timo Aila.
\newblock Analyzing and improving the image quality of stylegan.
\newblock In {\em Proceedings of the IEEE/CVF Conference on Computer Vision and
  Pattern Recognition}, pages 8110--8119, 2020.

\bibitem{CelebAMask-HQ}
Cheng-Han Lee, Ziwei Liu, Lingyun Wu, and Ping Luo.
\newblock Maskgan: Towards diverse and interactive facial image manipulation.
\newblock In {\em IEEE Conference on Computer Vision and Pattern Recognition
  (CVPR)}, 2020.

\bibitem{liao2020towards}
Yiyi Liao, Katja Schwarz, Lars Mescheder, and Andreas Geiger.
\newblock Towards unsupervised learning of generative models for 3d
  controllable image synthesis.
\newblock In {\em CVPR}, 2020.

\bibitem{lin2018st}
Chen-Hsuan Lin, Ersin Yumer, Oliver Wang, Eli Shechtman, and Simon Lucey.
\newblock St-gan: Spatial transformer generative adversarial networks for image
  compositing.
\newblock In {\em Proceedings of the IEEE Conference on Computer Vision and
  Pattern Recognition}, pages 9455--9464, 2018.

\bibitem{locatello2020object}
Francesco Locatello, Dirk Weissenborn, Thomas Unterthiner, Aravindh Mahendran,
  Georg Heigold, Jakob Uszkoreit, Alexey Dosovitskiy, and Thomas Kipf.
\newblock Object-centric learning with slot attention.
\newblock {\em Advances in Neural Information Processing Systems},
  33:11525--11538, 2020.

\bibitem{mildenhall2020nerf}
Ben Mildenhall, Pratul~P Srinivasan, Matthew Tancik, Jonathan~T Barron, Ravi
  Ramamoorthi, and Ren Ng.
\newblock Nerf: Representing scenes as neural radiance fields for view
  synthesis.
\newblock In {\em ECCV}, 2020.

\bibitem{nguyen2019hologan}
Thu Nguyen-Phuoc, Chuan Li, Lucas Theis, Christian Richardt, and Yong-Liang
  Yang.
\newblock Hologan: Unsupervised learning of 3d representations from natural
  images.
\newblock In {\em Proceedings of the IEEE/CVF International Conference on
  Computer Vision}, pages 7588--7597, 2019.

\bibitem{nguyen2020blockgan}
Thu Nguyen-Phuoc, Christian Richardt, Long Mai, Yong-Liang Yang, and Niloy
  Mitra.
\newblock Blockgan: Learning 3d object-aware scene representations from
  unlabelled images.
\newblock {\em arXiv preprint arXiv:2002.08988}, 2020.

\bibitem{Niemeyer2020GIRAFFE}
Michael Niemeyer and Andreas Geiger.
\newblock Giraffe: Representing scenes as compositional generative neural
  feature fields.
\newblock In {\em Proc. IEEE Conf. on Computer Vision and Pattern Recognition
  (CVPR)}, 2021.

\bibitem{or2021stylesdf}
Roy Or-El, Xuan Luo, Mengyi Shan, Eli Shechtman, Jeong~Joon Park, and Ira
  Kemelmacher-Shlizerman.
\newblock Stylesdf: High-resolution 3d-consistent image and geometry
  generation.
\newblock {\em arXiv e-prints}, pages arXiv--2112, 2021.

\bibitem{Ost_2021_CVPR}
Julian Ost, Fahim Mannan, Nils Thuerey, Julian Knodt, and Felix Heide.
\newblock Neural scene graphs for dynamic scenes.
\newblock In {\em Proceedings of the IEEE/CVF Conference on Computer Vision and
  Pattern Recognition (CVPR)}, pages 2856--2865, June 2021.

\bibitem{pan2021shadegan}
Xingang Pan, Xudong Xu, Chen~Change Loy, Christian Theobalt, and Bo Dai.
\newblock A shading-guided generative implicit model for shape-accurate
  3d-aware image synthesis.
\newblock In {\em Advances in Neural Information Processing Systems (NeurIPS)},
  2021.

\bibitem{richardson2021encoding}
Elad Richardson, Yuval Alaluf, Or Patashnik, Yotam Nitzan, Yaniv Azar, Stav
  Shapiro, and Daniel Cohen-Or.
\newblock Encoding in style: a stylegan encoder for image-to-image translation.
\newblock In {\em IEEE/CVF Conference on Computer Vision and Pattern
  Recognition (CVPR)}, June 2021.

\bibitem{saha2021LOHO}
Rohit Saha, Brendan Duke, Florian Shkurti, Graham Taylor, and Parham Aarabi.
\newblock Loho: Latent optimization of hairstyles via orthogonalization.
\newblock In {\em CVPR}, 2021.

\bibitem{Schwarz2020NEURIPS}
Katja Schwarz, Yiyi Liao, Michael Niemeyer, and Andreas Geiger.
\newblock Graf: Generative radiance fields for 3d-aware image synthesis.
\newblock In {\em Advances in Neural Information Processing Systems (NeurIPS)},
  2020.

\bibitem{stelzner2021decomposing}
Karl Stelzner, Kristian Kersting, and Adam~R Kosiorek.
\newblock Decomposing 3d scenes into objects via unsupervised volume
  segmentation.
\newblock {\em arXiv:2104.01148}, 2021.

\bibitem{sun2021fenerf}
Jingxiang Sun, Xuan Wang, Yong Zhang, Xiaoyu Li, Qi Zhang, Yebin Liu, and Jue
  Wang.
\newblock Fenerf: Face editing in neural radiance fields.
\newblock {\em arXiv preprint arXiv:2111.15490}, 2021.

\bibitem{szabo2019unsupervised}
Attila Szab{\'o}, Givi Meishvili, and Paolo Favaro.
\newblock Unsupervised generative 3d shape learning from natural images.
\newblock {\em arXiv preprint arXiv:1910.00287}, 2019.

\bibitem{Tan20}
Zhentao Tan, Menglei Chai, Dongdong Chen, Jing Liao, Qi Chu, Lu Yuan, Sergey
  Tulyakov, and Nenghai Yu.
\newblock Michigan: Multi-input-conditioned hair image generation for portrait
  editing.
\newblock {\em ACM Trans. Graph.}, 39(4), jul 2020.

\bibitem{tewari2020pie}
Ayush Tewari, Mohamed Elgharib, Florian Bernard, Hans-Peter Seidel, Patrick
  P{\'e}rez, Michael Zollh{\"o}fer, and Christian Theobalt.
\newblock Pie: Portrait image embedding for semantic control.
\newblock {\em ACM Transactions on Graphics (TOG)}, 39(6):1--14, 2020.

\bibitem{tsai2017deep}
Yi-Hsuan Tsai, Xiaohui Shen, Zhe Lin, Kalyan Sunkavalli, Xin Lu, and Ming-Hsuan
  Yang.
\newblock Deep image harmonization.
\newblock In {\em Proceedings of the IEEE Conference on Computer Vision and
  Pattern Recognition}, pages 3789--3797, 2017.

\bibitem{Wang21}
Ziyan Wang, Timur Bagautdinov, Stephen Lombardi, Tomas Simon, Jason Saragih,
  Jessica Hodgins, and Michael Zollhöfer.
\newblock Learning compositional radiance fields of dynamic human heads.
\newblock In {\em 2021 IEEE/CVF Conference on Computer Vision and Pattern
  Recognition (CVPR)}, pages 5700--5709, 2021.

\bibitem{wu2016learning}
Jiajun Wu, Chengkai Zhang, Tianfan Xue, Bill Freeman, and Josh Tenenbaum.
\newblock Learning a probabilistic latent space of object shapes via 3d
  generative-adversarial modeling.
\newblock In {\em NIPS}, 2016.

\bibitem{xia2021gan}
Weihao Xia, Yulun Zhang, Yujiu Yang, Jing-Hao Xue, Bolei Zhou, and Ming-Hsuan
  Yang.
\newblock Gan inversion: A survey.
\newblock {\em arXiv preprint arXiv:2101.05278}, 2021.

\bibitem{xu2021generative}
Xudong Xu, Xingang Pan, Dahua Lin, and Bo Dai.
\newblock Generative occupancy fields for 3d surface-aware image synthesis.
\newblock In {\em Advances in Neural Information Processing Systems(NeurIPS)},
  2021.

\bibitem{yang2021objectnerf}
Bangbang Yang, Yinda Zhang, Yinghao Xu, Yijin Li, Han Zhou, Hujun Bao, Guofeng
  Zhang, and Zhaopeng Cui.
\newblock Learning object-compositional neural radiance field for editable
  scene rendering.
\newblock In {\em International Conference on Computer Vision ({ICCV})},
  October 2021.

\bibitem{yenamandra2021i3dmm}
Tarun Yenamandra, Ayush Tewari, Florian Bernard, Hans-Peter Seidel, Mohamed
  Elgharib, Daniel Cremers, and Christian Theobalt.
\newblock i3dmm: Deep implicit 3d morphable model of human heads.
\newblock In {\em Proceedings of the IEEE/CVF Conference on Computer Vision and
  Pattern Recognition}, pages 12803--12813, 2021.

\bibitem{Yu-ECCV-BiSeNet-2018}
Changqian Yu, Jingbo Wang, Chao Peng, Changxin Gao, Gang Yu, and Nong Sang.
\newblock Bisenet: Bilateral segmentation network for real-time semantic
  segmentation.
\newblock In {\em European Conference on Computer Vision}, pages 334--349.
  Springer, 2018.

\bibitem{yu2021unsupervised}
Hong-Xing Yu, Leonidas~J Guibas, and Jiajun Wu.
\newblock Unsupervised discovery of object radiance fields.
\newblock {\em arXiv preprint arXiv:2107.07905}, 2021.

\bibitem{zhang2008cat}
Weiwei Zhang, Jian Sun, and Xiaoou Tang.
\newblock Cat head detection-how to effectively exploit shape and texture
  features.
\newblock In {\em European conference on computer vision}, pages 802--816.
  Springer, 2008.

\bibitem{zhang21}
Yuxuan Zhang, Huan Ling, Jun Gao, Kangxue Yin, Jean-Francois Lafleche, Adela
  Barriuso, Antonio Torralba, and Sanja Fidler.
\newblock Datasetgan: Efficient labeled data factory with minimal human effort.
\newblock In {\em CVPR}, 2021.

\bibitem{zhu2015learning}
Jun-Yan Zhu, Philipp Krahenbuhl, Eli Shechtman, and Alexei~A Efros.
\newblock Learning a discriminative model for the perception of realism in
  composite images.
\newblock In {\em Proceedings of the IEEE International Conference on Computer
  Vision}, pages 3943--3951, 2015.

\bibitem{Zhu21}
Peihao Zhu, Rameen Abdal, John Femiani, and Peter Wonka.
\newblock Barbershop: Gan-based image compositing using segmentation masks.
\newblock {\em ACM Trans. Graph.}, 40(6), dec 2021.

\end{thebibliography}
